\documentclass[12pt]{article}
\title{ Spin-dependent forces of quarks in baryon}
\author{Yu.A.Simonov\\
 Jefferson
Laboratory, Newport News, VA 23606, USA, and\\
 State Research
Center\\Institute of Theoretical and Experimental Physics, \\
Moscow, 117218 Russia}
 \date{}
\newcommand{\be}{\begin{equation}}
 \newcommand{\ee}{\end{equation}}

\def\fun#1#2{\lower3.6pt\vbox{\baselineskip0pt\lineskip.9pt
\ialign{$\mathsurround=0pt#1\hfil ##\hfil$\crcr#2\crcr\sim\crcr}}}

\newcommand{{\SD}}{\rm SD}

\newcommand{\ver}{\mbox{\boldmath${\rm r}$}}
\newcommand{\vesig}{\mbox{\boldmath${\rm \sigma}$}}

\newcommand{\vep}{\mbox{\boldmath${\rm p}$}}

\newcommand{\vez}{\mbox{\boldmath${\rm z}$}}

\newcommand{\veL}{\mbox{\boldmath${\rm L}$}}
\newcommand{\veR}{\mbox{\boldmath${\rm R}$}}

\newcommand{\ven}{\mbox{\boldmath${\rm n}$}}
\newcommand{\veu}{\mbox{\boldmath${\rm u}$}}

\newcommand{\veB}{\mbox{\boldmath${\rm B}$}}

\newcommand{\veE}{\mbox{\boldmath${\rm E}$}}

\newcommand{\veal}{\mbox{\boldmath${\rm \alpha}$}}

\newcommand{\llan}{\langle\langle}
\newcommand{\rran}{\rangle\rangle}
\newcommand{\lan}{\langle}
\newcommand{\ran}{\rangle}
\begin{document}
\maketitle

\begin{abstract}
Nonperturbative spin-dependent forces of quarks in a baryon are
calculated  directly from the QCD Lagrangian in the framework of
the Field Correlator Method both for heavy and light quarks.
Resulting forces contain terms of 5 different structures,only one
being known before in asymptotic form.Perturbative terms obtained
by the same method are standard and have different signs and structures
with respect to the corresponding nonperturbative ones,implying
possible cancellations for some baryonic states.
\end{abstract}

\section{Introduction}
The spin structure of baryons presents a still unsolved problem,both on
partonic and quark model level. For excited baryon spectrum the
apparent small spin-orbit splitting of some baryonic states is a
topic of vivid discussions \cite{1}-\cite{3}. Some baryonic
states, like Roper resonance $N(1440)$ or $\Lambda(1405)$ are not
yet explained in the traditional framework of relativistic quark
model (RQM) \cite{2,4}.More detailed information about the spin
structure of baryons comes from the polarization experiments on
electroproduction of excited resonances, \cite{5,6} which
effectively measure the convolution of the  baryon wave function,
and is very sensitive to its structure.

Meanwhile the theoretical knowledge of the quark spin forces in
the baryon is limited to the perturbative expressions calculated
decades  ago  \cite{7}, and nonperturbative spin-orbit Thomas
term, written in the framework of
RQM \cite{8}.

In applying these results to light baryons the notion of
constituent quark masses is introduced in  RQM,these appear in
spin-dependent (SD) forces and play role of fitting parameters.

It is the purpose of the present paper to derive SD forces in a
baryon in a most straightforward way from the QCD Lagrangian with
the nonperturbative vacuum described by vacuum field correlators
\cite{9,10}. Limiting to the lowest (Gaussian) field correlators
one can express all terms of SD forces through the scalar
functions $D$ and $D_1$ representing this Gaussian correlator
\cite{11}. High accuracy of such procedure is supported by recent
lattice data \cite{12} and contribution of higher correlators can
be estimated to be of the order of few percent \cite{13}.

The functions $D,D_1$ are  themselves measured on the lattice
\cite{14} and also  found in analytic approaches \cite{15, 16}.

An essential element of the present approach is that it is not
connected to the heavy mass expansion, and can be applied also to
light quarks in a  baryon. In this case an effective Hamiltonian
is constructed from  first principles, which contains einbein
(auxiliary) fields. It was shown previously, that stationary point
of these einbein fields yields exactly the constituent quark masses
which can be expressed unambigiously through the only parameter of
this approach -- the string tension $\sigma$.

The decisive check of this procedure is the calculation of baryon
magnetic moments, since they  are inversely proportional to the
quark constituent masses. That was done in \cite{17} and results
agreed with all known experimental data within $\sim 10\%$.

The SD forces derived below in the paper are computed as a series
in field correlators (cumulants) with growing powers of fields.

The lowest (Gaussian) term yields SD forces inversely proportional
to the square of constituent quark masses. Having in mind the high
accuracy of the Gaussian approximation \cite{13} and baryon magnetic
moments \cite{17} one should expect that SD forces found below
have the accuracy of the order of 10\%.

Analogous expressions for heavy quarkonia \cite{18,19} and light
mesons \cite{20} have been reported earlier and estimated for
realistic meson system respectively in \cite{21} and \cite{22}.

The plan of the paper is as follows. In chapter 2 a general
expression for the 3q baryon Green's function is introduced, and
the Fock-Feynman-Schwinger (world-line) formalism is used to
reveal the dependence on gauge fields with spin operators
explicitly written. Averaging over those with the help of the
Field Correlator Method (FCM) one finally obtains an expression
for the Green's function written in terms of field correlators.

In chapter 3 a special case of heavy quark masses is considered
and all SD forces are obtained in closed form, expressed in terms
of correlator functions $D$ and $D_1$.

In chapter 4 the perturbative contribution to SD forces is written
down. In chapter 5 a general case is considered when current quark
masses can also be vanishingly small, and SD forces are written
again in terms of integrals over functions $D$ and $D_1$ with
constituent (dynamical) masses entering in the denominator.

Chapter 6 is devoted to the discussion of the relativistic
structure of SD forces in the excited baryon spectrum. Comparison
to other results in the literature is also made and possible
extension of the method is suggested. Main points of
the paper are summarized in Conclusions.

\section{$3q$ Green's
function with spin insertions}

Following \cite{9,10,23} we consider the $3q$ Green's function,
which can be written as
\be
G_{3q} (x,y) =tr_L \left[ \Gamma_{out} \prod^3_{i=1} (m_i-\hat D)
\int D\mu_i Dz^{(i)} e^{-K_i}\lan W_3\exp g \sigma F\ran
\Gamma_{in}\right]\label{1} \ee where $tr_L$ is the trace over
Dirac matrix indices, $\Gamma_{out} (\Gamma_{in})$ are final
(initial) state operators  creating given $J^{PC}$ assignment to
the $3q$ state and we have also denoted as in \cite{17, 20}
\be
K_i=\int^T_0 dt \left [\frac{m^2_i}{2\mu_i(t)} +\frac{\mu_i(t)}{2}
(\dot{\vez}^2(t) +1)\right],\label{2}\ee
\be
\lan W_3\exp (g\sigma F)\ran =tr_Y\exp\left[ \sum^\infty_{n=0}
\frac{(ig)^n}{n!}\int \llan F(1)... F(n)\rran d\rho (1).. d\rho
(n)\right].\label{3}\ee Here and in what follows $F(1)$ is always
implied to be gauge-transported to one point $x_0$,namely
$F(1) \equiv F(z(1),x_0)=\Phi(x_0,z(1))F(z(1))\Phi(z(1),x_0)$
where $\Phi(x,y)$ is defined in the Appendix,
and finally, $d\rho(n)=\sum^3_{i=1}d\rho^{(i)} (n)$, with
\be
d\rho^{(i)} (n) \equiv ds^{(i)}_{\mu_n\nu_n}
(u^{(n)})+\frac{1}{i} \sigma^{(i)}_{\mu_n\nu_n}
\frac{dt_n}{2\mu_i(t_n)}.\label{4}\ee

The integration in (\ref{3}) extends over all 3 lobes of the
minimal area surface $(S_1+S_2+S_3)$ inside the quark trajectories
$z^{(i)}(t)$  and the string-junction trajectory $z^{(Y)}(t)$. We
have also denoted $\sigma_{\mu\nu} =\frac{1}{4i}
(\gamma_\mu\gamma_\nu-\gamma_\nu\gamma_\mu)$ and everywhere
Euclidean space-time is used (till the last moment when the
resulting Hamiltonian is obtained in Minkowski space-time) with
$\gamma$-matrices $$\gamma_4=\gamma_0\equiv \beta;~~
\gamma_i=-i\beta\alpha_i,~~
\gamma_\mu\gamma_\nu+\gamma_\nu\gamma_\mu=2\delta_{\mu\nu}.$$

Note also that notation $tr_Y$ means
\be
tr_YP\equiv \frac16 e_{abc} e_{a'b'c'} P_{abc/a'b'c'}.\label{5}\ee

 In the combination $F(k) d\rho(k)$ in (\ref{3}) one can
 write
 \be
 F_{\mu\nu}\sigma^{(i)}_{\mu\nu} =\left(\begin{array}{ll}
\vesig^{(i)} \veB,&\vesig^{(i)}\veE\\ \vesig^{(i)} \veE,&
\vesig^{(i)} \veB\end{array}\right),~~ i=1,2,3.\label{6} \ee As it
was shown in \cite{10} the spin-independent part of (\ref{3})
which obtains neglecting the $\Sigma$ term in (\ref{4}),
 yields at large quark separations, $|\vez^{(i)}-\vez^{(Y)}|\gg
 T_g,~~ i=1,2,3,$ the familiar area-law asymptotics
 \be
 \lan W_3\ran=\exp [ -\sigma (S_1+S_2+S_3)]\label{7}
 \ee
 implying linear confinement for each quark.
 In what follows we shall use the general expression (1) to
 derive the spin-dependent part of interaction both for heavy quarks
 (expansion in inverse powers of mass) and for light quarks.

 \section{Spin-dependent interaction in $1/m$ expansion}

 To illustrate the method we shall start with the derivation of
 spin-dependent (SD) forces via $1/m$ expansion. Defining the SD
 potential as $V_{{\SD}}$, one can write $G_{3q}\sim e^{-TV_{{\SD}}}\sim
 1- TV_{{\SD}}$, and for $V_{{\SD}} $
the following general form will be obtained below, similar (but
 not identical) to the corresponding form for heavy quarkonia
 \cite{24,18},
 $$
 V_{{\SD}} (\veR^{(1)},\veR^{(2)}, \veR^{(3)})
 =\sum^3_{i=1}\frac{\vesig^{(i)}\veL^{(i)}}{2m^2_i}\left
 (\frac{1}{R^{(i)}}\frac{dV_1}{dR^{(i)}}
 +\frac{1}{2R^{(i)}}\frac{d\varepsilon}{dR^{(i)}}\right)
 $$
 $$
+ \frac{1}{N_c-1}\sum_{i<j}\frac{(\vesig^{(i)}\veL^{(j)}
 +\vesig^{(j)}\veL^{(i)})}{2m_im_j}\frac{1}{R^{(j)}}
\frac{dV_2(R^{(i)}, R^{(j)})}{dR^{(j)}}+$$
\be
+\sum_{i<j}\left [
\frac{(\vesig^{(i)}\vesig^{(j)})V_4(R_{ij})}{12m_im_j(N_c-1)}+
\frac{3(\vesig^{(i)}\ven)(\vesig^{(j)}\ven)-(\vesig^{(i)}\vesig^{(j)})}
{12 m_im_j(N_c-1)}
V_3(R_{ij})\right ]+V_5,\label{8} \ee
where $\ven =\frac{\veR_{ij}}{R_{ij}}$,$\veR_{ij}=\veR_i-\veR_j$.
We assume that current quark masses are large, $m_i\gg
\sqrt{\sigma}, ~~ i=1,2,3,$ and hence also $\mu_i$ are large,
since the latter are defined through $m_i$ and  $\sigma$ in the
stationary point analysis \cite{10,17} and always satisfy $\mu_i=
m_i+ O(1/m_i)$. Hence for simplicity we keep in the following
$\mu_i=m_i\gg \sqrt{\sigma}$ and expand in inverse powers of
$1/m_i$.

As it was observed in \cite{18,19} the SD terms of the lowest
order $(1/m^2_i,~~\frac{1}{m_im_j})$ come from 3 different
sources:

A) Diagonal terms in (\ref{6}) are kept together with diagonal
terms in $\Lambda_i\equiv (m_i+\frac12\gamma_\mu\dot z_\mu)$,
yielding one power of $1/m_i$. An additional power of
$\frac{1}{m_i}$ or $\frac{1}{m_j} $ then comes from the expansion
of $\lan W_3\ran$. This yields spin-orbit terms $V'_1, V'_2$ and
$V_5$.

B) The off-diagonal terms are kept both in (\ref{6}) and in
$\Lambda_i$. This gives spin-orbit potential $\frac{d\varepsilon}{dR}$.

C) Diagonal terms from two matrices (\ref{6}) with $i\neq j$ are
retained. This yields spin-spin potentials $V_3$ and $V_4$. We now
calculate the SD contributions from A) -- C) point by point.

A) From (\ref{3}), (\ref{4}) one gets for $i=1$ \be \lan
tr_Y[(1+\frac{g}{2m_1} \sigma_k^{(1)} \int^T_0 B_k(z^{(1)},
t^{(1)}) dt_1)W_3]\ran \approx 1 -TV_{\SD}^{(1)} \label{9}\ee
Using the relation $ig F_{\mu\nu} W=
\frac{\delta}{\delta\sigma_{\mu\nu}(z)} W$ which obtains easily
with nonabelian Stokes representation for $W$, one has
\be
\lan tr_YF_{\lambda\sigma}(x,z_0) W(C)\ran =tr_Y\{ ig \int
ds_{\mu\nu}(z) F_{\mu\nu} (z,z_0) F_{\lambda\sigma}(x,z_0)
W_3(C)\}\label{10}\ee one can rewrite the l.h.s. of (\ref{9}) as
\be
tr_YW+\frac{ig^2}{2m_1} \sigma_k^{(1)} \int^T_0 dt_1 \lan tr_Y B_k
(z^{(1)}, t^{(1)} ) \int_{S(C')} ds_{\mu\nu} (u) F_{\mu\nu}
(u,x_0)  W_3(C')\ran.\label{11}\ee

In (\ref{10}) and (\ref{11}) the common reference point $z_0$ is
chosen to make both expressions gauge invariant; as it will be
seen, this point will not appear in the final equations.

In (\ref{11}) the contour $C'$ is deformed due to orbital momentum
of quarks as compared to the zeroth-order contour $C_0$ consisting
of straight lines. This is essential since otherwise the vacuum
average  $\lan B_k W_3(C_0)\ran$ vanishes since it is odd with
respect to reflection $z_i\to -z_{i}$ $i\neq k$. Therefore all
nonzero contribution in (11) is due to deflection of the quark
path in $C'$ from the straight line in $C_0$.

At this point we shall describe the quark trajectory
$z_\mu^{(i)}(t) $ and the corresponding string piece $W_\mu^{(i)}$
from  the quark position to the string junction (which we for
simplicity take at the origin).
\be
w_\mu^{(i)} (t,\beta) = z^{(i)}_\mu(t)\beta, ~~  1\geq \beta \geq
0 \label{12}\ee
\be
ds_{ik}^{(i)} = d\beta ^{(i)} dt e_{ikm} \frac{\beta
L_m^{(i)}}{im_i} \label{13}\ee where $L^{(i)}_m$ is the
(Minkowskian) angular momentum of the $i$-th quark \be L^{(i)}_s =
im_i e_{skm} R^{(i)}_k \dot{z}_m^{(i)},~~ \veR^{(i)} =
\vez^{(i)}-\vez^{(Y)} =\vez^{(i)}.\label{14}\ee

 Similarly
$d\sigma^{(i)}_{k4} = R^{(i)}_k d\beta^{(i)} du_4$,
 and one arrives at the  result
 $$
 \lan B_k(z^{(i)}, t_i) W_3(C')\ran =ig \int d\beta^{(i)} du_4
 \frac{\beta^{(i)}L^{(i)}_n}{im_i} \lan B_k B_n (u_4,\beta) W\ran+
 $$
\be
+ig \int d\beta^{(i)}  du_4  R^{(i)}_l(u_4) \lan B_kE_l
(u_4,\beta^{(i)}) W\ran . \label{15}\ee Denoting \be \lan
B_k(z^{(1)}, t_1) E_i (\veu, u_4) W_3\ran \equiv e_{kin}
(u_n-z_n^{(1)}) \frac{\partial\Lambda_0}{\partial u_4}\label{16}
\ee one obtains $$ \vesig^{(i)}\veL^{(i)}\left (
\frac{1}{R}\frac{d V_1}{dR}\right)^{(i)} =- g^2\int^1_0 \beta
d\beta \int^T_0 du_4 \frac{\sigma_k^{(i)} L_n^{(i)}}{\lan
W_3\ran}\lan B_k(z^{(i)}, t_i)B_n (\beta z^{i}, u_4) $$
\be
-\frac{\vesig^{(i)}\veL^{(i)}}{\lan W_3\ran} \int^T_0 du_4
(u_4-t_1) d\beta \frac{\partial\Lambda_0}{\partial u_4}\label{17}
\ee where we have used the relation $u_n(u_4) -z_n^{(1)} (t_1)
\cong \dot z_n^{(1)} (u_4) (u_4-t_1)$.

 Until now we have not used
the Gaussian dominance of the vacuum, i.e. the fact that $\lan
W_3\ran$ is saturated  by the lowest cumulant $\lan FF\ran$, which
is found to be an accurate approximation \cite{13}. Using it one
can write $\lan FF W_3\ran \to \lan FF\ran \lan W_3\ran$, and
introduce scalar functions $D,D_1$ for tensor $\lan FF\ran$, as it
was done in \cite{11}.

Referring the reader to the Appendix for the corresponding
relations, one finally obtains $$\left ( \frac{1}{R}
\frac{dV_1}{dR} \right)^{(i)} =- \int^R_0 \frac{d\lambda}{R}
(1-\frac{\lambda}{R})\int^\infty_{-\infty} d\nu [ D(\lambda, \nu)
+ D_1(\lambda, \nu) + \lambda^2\frac{\partial D_1}{\partial
\lambda^2}]- $$
\be
-\int^\infty_{-\infty} \nu^2 d\nu \int^R_0 \frac{d\lambda}{R}
\frac{\partial D_1}{\partial \lambda^2}.\label{18} \ee Till now we
have taken into account the interaction of the spin of the $i$-th
quark with the surface $S_i$, which yields the term $(\frac{1}{R}
V'_1)^{(i)}$ multiplied with $(m^2_i)^{-1}$. At this point we
consider the interaction of the $i$-th quark spin with the
(deformed) surface $S_j$, which will give the term $V'_2$ in
(\ref{8}). For this one needs to consider a vacuum average of two
$F$'s from two different surfaces $S_i$ and $S_j$.

In general we have for two $F$'s transported to the same point $x$
$(\alpha,... \eta$ are fundamental color indices)
\be
\lan F (u,x)_{\alpha\xi} F(v,x)_{\gamma\eta} \ran =\frac{\lan tr
FF\ran}{N^2_c-1} (\delta_{\alpha\eta} \delta_{\xi\gamma}
-\frac{1}{N_c} \delta_{\alpha\xi} \delta_{\gamma\eta}).\label{19}
\ee Taking into account (\ref{5}) and the relation
\be
tr_Y \Phi_{\alpha\alpha'}(x,y) \Phi_{\beta\beta'} (x,y)
\Phi_{\gamma\gamma'} (x,y) \equiv 1
\label{20} \ee one obtains
\be
tr_Y\lan F_{\alpha\alpha'} (u,x) F_{\beta\beta'} (v, x) \ran =
\frac{\lan tr F(u,x)F(v,x)\ran}{N_c(N_c-1)}\label{21}\ee where we
 have also accounted for different orientation of plaquettes in
 $S_i$ and $S_j$.

Now proceeding as in (\ref{15}) one has $$\lan B_k^{(i)} W_3
(C'_j)\ran = g\int^1_0 \beta^{(j)} d\beta^{(j)} du_4^{(j)}
\frac{L_n^{(j)}}{m_j} \lan B_k^{(i)} B_n^{(j)}
(u_4^{(j)},\beta^{(j)} R^{(j)}) W_3\ran + $$
\be
+ig \int d\beta^{(j)} du_4^{(j)} R^{(j)}_l (u^{(j)}_4) \lan
B_k^{(i)} E_l^{(j)} W_3\ran.\label{22} \ee
 At this point one can use Gaussian dominance and relations
 (\ref{21}) to obtain finally
 \be
 \left (\frac{1}{R}V_2' (R)\right )_j =  \int^1_0 d\beta^{(j)}
 \beta^{(j)} \int^\infty_{-\infty} d\nu [ D(r^{(ij)}, \nu) + D_1+
 ((r^{(ij)})^2 +\nu^2) \frac{\partial
 D_1}{\partial(r^{(ij)})^2}].\label{23}\ee

 In a similar way one obtains from the first term on the r.h.s.
 of (\ref{22}),with the use of the last term on  the r.h.s. of
 (A1).
 \be
 V_5 =-\sum_{i>j} \int \frac{(\vesig^{(i)}
 \ver^{(ij)})(\veL^{(j)}\ver^{(ij)})}{2m_im_j(N_c-1)} \beta^{(j)}
 d\beta^{(j)} d\nu \frac{\partial
 D_1(r^{(ij)},\nu)}{\partial(r^{(ij)})^2}.\label{24}\ee

 Here we have defined
 $\ver^{(ij)}= \veR^{(i)} -\beta^{(j)} \veR^{(j)}$, and one should
 take into account, that $D, D_1$ depend on their arguments as
 $$D(r,\nu)= D(\sqrt{r^2+\nu^2}).$$

 This concludes derivation of terms  with the procedure A) and one
 goes over to the next point.

 B) Following (\ref{9}) one can write for the corresponding term
 of a given quark ($i$)
 \be
 V_{SD}^{(\varepsilon)} T=-\frac{g}{(2m_i)^2} \lan \left (
 \begin{array}{ll}
 m_i+\mu_i&-\vesig^{(i)}\vep^{(i)}\\
 \vesig^{(i)} \vep^{(i)} & m_i-\mu_i\end{array}\right)\left(
\begin{array}{ll}
 0&\vesig^{(i)}\veE\\
 \vesig^{(i)} \veE & 0\end{array}\right) W_3(C)\ran.\label{25}
 \ee
 Now one can use relation
 \be
 \lan E_k(z^{(i)}, t^{(i)}) W_3(C_0)\ran =\frac{\delta\lan
 W_3(C_0)\ran}{ig\delta \sigma_{k4} (z^{(i)},
 t^{(i)})}=-\frac{1}{ig} \frac{\partial\varepsilon}{\partial
 R^{(i)}_k}\lan W_3(C_0)\ran\label{26}\ee
 where the following notation was introduced for the
 spin-independent potential $\varepsilon (R^{(1)}, R^{(2)},
 R^{(3)})$.
\be
\lan W_3(C_0)\ran =\exp
[-\varepsilon(R^{(1)},R^{(2)},R^{(3)})T].\label{27}\ee

Note that in (\ref{25}-\ref{27}) one can keep in  $W_3(C)$ the
unperturbed (straight-line) contours for quark  trajectories since
the  prefactor in (\ref{25}) is already  $O(1/m^2)$. Keeping in
mind relation \be \sigma_k^{(i)} E_k^{(i)}
\sigma^{(i)}_lp^{(i)}_l= E^{(i)}_k p^{(i)}_k+i e_{kln}
\sigma_n^{(i)} E_k^{(i)} p^{(i)}_l\label{28}\ee one recovers the
second term on the r.h.s. of (\ref{8}).

C) Here one considers spin-spin interaction and the corresponding
term looks like $$ 1-V_{\SD} T= tr_Y\sum_{i>j} \lan
(1+\frac{g}{2m_i} \int\vesig^{(i)} \veB^{(i)} (z^{(i)}, t_i)
dt_i)\times $$
\be
\times (1+\frac{g}{2m_j}\int \vesig ^{(j)} \veB^{(j)} (z^{(j)},
t_j) dt_j) W_3\ran \label{29}\ee Identifying spin-spin terms in
(\ref{29}) and using (\ref{21}) and relations for $\lan BBW_3\ran$
in Appendix one arrives at $$
V_{\SD}^{(\sigma\sigma)}=\sum_{i<j}\int^\infty_{-\infty}
\frac{d\nu \sigma_k^{(i)} \sigma^{(j)}_{k'}}
{4m_im_j(N_c-1)}[\delta_{kk'} (D+D_1+(\veu)^2\frac{\partial
D_1}{\partial(\veu)^2})- $$
\be
-u_k u_{k'}\frac{\partial D_1}{\partial(u)^2}]\label{30}\ee
where notations are used
\be
\veu = \veR^{(i)}-\veR^{(j)},~~ \nu=t_i-t_j.\label{31}\ee
Rewriting (\ref{30}) as
 \be V_{\SD}^{(\sigma\sigma)}=\sum_{i<j}
\frac{\vesig^{(i)}\vesig^{(j)} V_4(u)+ S_{ij}
V_3(u)}{12 m_im_j(N_c-1)}\label{32}\ee one has
\be
V_4(u) =\int^\infty_{-\infty} d\nu (3 D(u,\nu) +3D_1 (u,\nu)
+2u^2\frac{\partial D_1}{\partial u^2})\label{33}\ee
\be
V_3(u) =-\int^\infty_{-\infty} d\nu u^2\frac{\partial
D_1(u,\nu)}{\partial u^2},\label{34} \ee
\be
S_{ij}=3(\vesig^{(i)} \ven) (\vesig^{(j)} \ven)-
\vesig^{(i)}\vesig^{(j)},~~ \ven=
\frac{\veu}{|\veu|}.\label{35}\ee

This concludes definition of all NP spin-dependent terms in
(\ref{8}) to the order $O(1/m^2)$ and in the approximation when
only lowest, $\lan FF\ran $, correlator is retained in the Wilson
loop.

Now comparing our expressions for $V'_1, V'_2, \varepsilon',
V_3,V_4$ with the corresponding ones for heavy $Q\bar Q$ case,
given in \cite{18,19}, one can see  that they coincide exactly,
the only difference being that one should sum up over all 3 quarks
for $V'$ and $\varepsilon'$, and take a double sum, $i<j$, for
$V'_2, V_3,V_4$. In addition there is a term $V_5$  which is of
3 body character and vanishes in two-body situation since
in that situation $\veL^{(j)}\ver^{(ij)}=\veL^{(j)} \ver\equiv 0.$

\section{Perturbative spin-dependent forces}

We are now in position to consider also perturbative contributions
to the SD potentials, which were calculated in \cite{7}. The
easiest way for us is to remember, that to the lowest order,
$O(\alpha_s)$, all perturbative terms are pair-wise interaction of
quarks, and they can be reconstructed from the expressions
obtained above, Eqs. (\ref{18}), (\ref{23}),
(\ref{24}),(\ref{32})-(\ref{34}), using $O(\alpha_s)$ contribution
to $D_1(x)$, while $D(x)$ does not have contributions at this
order, \cite{25}
\be
D(x)=D^{NP}(x), ~~D_1(x)=\frac{16\alpha_s}{3\pi x^4}+ D_1^{NP}
(x). \label{36}\ee It is rewarding to realize that $D_1$ does not
enter into $V'_1$ (terms containing $D_1$ in (\ref{18}) cancel
exactly), so that perturbative contribution occurs only in the
nondiagonal, $i\neq j$, terms in (\ref{8}) and in $\varepsilon'$.

In the $Q\bar Q$ case analogous calculations have been done and
compared to standard ones in \cite{18,19}.

We start with the $\frac{d\varepsilon}{dR}$ term and rewrite it in the
original form (\ref{26}), not assuming that $\varepsilon $ depends
on $R^{(i)}$ only, but also on $\veR^{(i)}-\veR^{(j)}$, as it is for
the Coulomb term to be added to $\varepsilon$. In this way one
obtains from (\ref{25}), (\ref{26}),(\ref{28})
\be
V_{\SD}^{(\varepsilon)}
=\frac{e_{kln}p_l^{(i)}\sigma^{(i)}_n}{4m^2_i} \frac{\partial
\varepsilon}{\partial R_k^{(i)}}\label{37} \ee and  for
$\varepsilon \to \varepsilon + V_{coul},~~
V_{coul}=-\frac{2\alpha_s}{3}\sum_{i>j}
\frac{1}{|R^{(i)}-R^{(j)}|}$ one obtains
\be
V_{\SD}^{(\varepsilon,pert)}= \frac{2\alpha_s}{3} \sum_{i>j}
\left[
\frac{(\veR^{(ij)}\times\vep^{(i)})\vesig^{(i)}}{4m^2_i(R^{(ij)})^3}+
\frac{(\veR^{(ji)}\times\vep^{(j)})\vesig^{(j)}}{4m^2_j(R^{(ij)})^3}\right].\label{38}\ee

This expression coincides with the corresponding in \cite{7}.
Consider now nondiagonal spin-orbit term, equivalent to $V'_2$.
Instead of using replacement (\ref{36}) and doing integrations in
(\ref{23}), we start from more general expression (\ref{9}) to
derive
\be
V_{\SD}^{(2,pert)}=- \frac{2\alpha_s}{3(N_c-1)} \sum_{i>j}
\frac{[\vesig^{(i)}(\veR^{(ij)} \times
\vep^{(j)})+\vesig^{(j)}(\veR^{(ji)} \times\vep^{(i)})]}{
m_im_j(R^{(ij)})^3}\label{39}\ee which is again in agreement with
\cite{7}.

 Next we consider the spin-spin interaction. Here it is
straightforward to replace in $V_4$ (\ref{33}) $D_1$ as in
(\ref{36}) $D_1\to D_1^{(pert)}=\frac{16\alpha_s}{3\pi x^4}$ to
obtain $$ V_4^{(pert)} (r) =\int^\infty_{-\infty} d\nu
(3D_1^{(pert)}(r,\nu)+2r^2\frac{\partial D_1^{(pert)}}{\partial
r^2})=$$
\be
=-\frac{8\alpha_s}{3r} \left(\frac{\partial^2}{\partial
r^2}+\frac{2}{r} \frac{\partial}{\partial r}\right)
\frac{1}{r}=\frac{32 \pi \alpha_s}{3} \delta^{(3)}
(\ver).\label{40}\ee In a similar way one obtains for
$V_3^{(pert)}$,
\be
V_3^{(pert)}(r) = \frac{4\alpha_s}{r^3}.\label{41}\ee

One can also persuade oneself that $V_5$ has no pertubative
counterpart  since there  $\veL^{(j)} \to \veL^{(ij)}$ and it is
orthogonal to $\ver^{(ij)}$  and therefore the total perturbative
SD contribution to the order $O(\alpha_s)$ can be written as
\be
V_{\SD}^{(pert)} =V_{\SD}^{(\varepsilon, pert)}+
V_{\SD}^{(2,pert)}+ \sum_{i<j} \frac{\vesig^{(i)}\vesig^{(j)}
V_4^{(pert)}(R^{(ij)})+ S_{ij} V_3^{(pert)}(R^{(ij)})}{12
m_im_j(N_c-1)}\label{42}\ee where explicit form of 4 terms on the
r.h.s. of (\ref{42}) is given in (\ref{38}), (\ref{39}),
(\ref{40}) and (\ref{41}).

Our results for $ V_{\SD}^{(2,pert)}, V_3^{(pert)}, V_4^{(pert)}$
coincide with the corresponding expressions in \cite{8}, however
our $V_{\SD}^{(\varepsilon,pert)}$ is 2 times smaller than the
corresponding term in \cite{8}. In the next sections we shall
argue that for light quark  this term gets indeed twice as big,
since there one should replace $m_i\to \mu_i$, and  for
$V_{SD}^{(\varepsilon)}$ the coefficient appears to be  2 times
larger.

\section{Spin-dependent forces for light quarks}

In sections 3, 4 the SD forces have been obtained as an expansion in
$1/m_i, 1/m_j$ taking all 3 quark current masses large, $m_i\gg
\sqrt{\sigma},~~ i=1,2,3.$

It was noticed before \cite{20} however, that general expressions
(\ref{1})-(\ref{4}) for Green's functions written in FSR, with the
einbein function $\mu_i(t)$ introduced as in \cite{10}, allow to
obtain expressions for SD forces also for light quarks without
$1/m$  expansion, and the corresponding terms for the meson case
have been written before \cite{10,22}. Below we demonstrate in
this section that the same procedure works also for the $3q$ case
with  light current quarks masses as well.

We start again with general form (\ref{3}) and instead of
expansion in $1/m$ (or $1/\mu$ which is equivalent for heavy
quark) shall do the only approximation: keeping in the sum in the
exponent (\ref{3}) the lowest (Gaussian) cumulant  $\llan F(1)
F(2)\rran$. This approximation was recently supported by lattice
data for Casimir scaling \cite{12}, while higher cumulants provide
(for Wilson loop) less than 2\% \cite{13}.

We start with the spin-spin interaction, which is easily obtained
keeping in (\ref{3}) the bilocal term and in (\ref{4}) only the
$\sigma$- dependent term. In (\ref{3}) one should take into
account that $F(1)$ and $F(2)$ belong to different lopes $S_1,
S_2$ of the $S_{123}$ surface, hence $n=1$ for each of them;
moreover one uses (\ref{19}) and (\ref{21}), the latter with
opposite sign, since orientation of $S(1,3)$ and $S(2,3)$ is the
same in our case. As a result one obtains
\be
V_{\SD}^{(\sigma\sigma)} =\sum_{i>j} \int^\infty_{-\infty}
\frac{d(t_i-t_j)}{4\mu_i\mu_j(N_c-1)} \sigma^{(i)}_{\mu_i\nu_i}
\sigma^{(j)}_{\mu_j\nu_j} \lan F_{\mu_i\nu_i} (i) F_{\mu_j\nu_j}
(j)\ran.\label{43}\ee The combination $\sigma^{(i)}\sigma^{(j)}$
in (\ref{43}) is a product of two $4\times 4$ matrices, which can
be split into the product of Pauli spin matrices $\sigma_i$ and
chiral $2\times 2$ matrices $\hat 1$ and $\hat \rho_1\equiv \left
(\begin{array}{ll} 0&1\\ 1&0\end{array}\right).$  Thus one can
rewrite (\ref{43}) as
 $$ V_{\SD}^{(\sigma\sigma)} =\sum_{i>j}
\int^\infty_{-\infty} \frac{g^2d(t_i-t_j)}{4\mu_i\mu_j(N_c-1)}
\sigma_m^{(i)}\sigma_n^{(j)}[\lan tr B_m(i) B_n(j)\ran(\hat
1\times \hat 1)+ $$ $$ + \lan tr E_m(i) E_n(j)\ran(\hat
\rho_1\times \hat \rho_1)+ \lan tr B_m(i) E_n(j)\ran(\hat 1\times
\hat\rho_ 1)+ $$
 $$\lan tr E_m(i) B_n(j)\ran(\hat\rho_1
\times \hat 1)]\equiv  V_{\SD}^{(\sigma\sigma)}(BB)
+V_{\SD}^{(\sigma\sigma)}(EE)+ $$
\be
 V_{\SD}^{(\sigma\sigma)}(BE)
+V_{\SD}^{(\sigma\sigma)}(EB).\label{44} \ee Using formulas from
Appendix for correlators of $B,E$ one has \be
V_{\SD}^{(\sigma\sigma)}(B_1B) =\sum_{i>j}
\frac{\vesig^{(i)}\vesig^{(j)} V_4(u)+ S_{ij}
V_3(u)}{12 \mu_i\mu_j(N_c-1)}(\hat 1\times \hat 1)
\label{45}\ee where $\veu \equiv \veR^{(i)} - \veR^{(j)}$.

One can see that (\ref{45}) coincides with (\ref{32}-\ref{35})
with substitution $m_i, m_j\to \mu_i,\mu_j$. For $EE$  term one
obtains
 \be
V_{\SD}^{(\sigma\sigma)}(EE) =\sum_{i>j}
\frac{\vesig^{(i)}\vesig^{(j)} \tilde V_4(u)+ S_{ij} \tilde
V_3(u)}{12 \mu_i\mu_j(N_C-1)}(\hat \rho_1\times \hat
\rho_1) \label{46}\ee where we have defined
\be
\tilde V_4(u) =\int^\infty_{-\infty} d\nu (3 D(u,\nu) +3D_1(u,\nu)
+(3\nu^2+u^2) \frac{\partial D_1(u,\nu)}{\partial
\nu^2})\label{47}\ee
\be
\tilde V_3(u) =\int^\infty_{-\infty} d\nu u^2  \frac{\partial
D_1(u,\nu)}{\partial u^2}=-V_3(u).\label{48}\ee

Finally for the last two terms in (\ref{44}) one has
 \be
V_{\SD}^{(\sigma\sigma)}(BE)=- V_{\SD}^{(\sigma\sigma)}(EB)=
\sum_{i>j} \frac{(\vesig^{(i)}\times \vesig^{(j)})\frac{1}{2i}
(\frac{\vep^{(i)}}{\mu_i}+\frac{\vep^{(j)}}{\mu_j})} {4
\mu_i\mu_j(N_c-1)} \int^\infty_{-\infty} \frac{\partial
D_1(u,\nu)}{\partial u^2}\nu^2 d\nu.\label{49}\ee
 This concludes
calculation of spin-spin interaction.

We turn now to the calculation of spin-orbit terms. The
corresponding expression in (\ref{3}) can be written as $$ \lan
W_3 \exp (g\sigma F)\ran_{so}= \exp \left\{ \sum^3_{i=1} i \int
\frac{dt_i}{2\mu_i} \sigma^{(i)}_{\mu\nu}
ds^{(i)}_{\rho\sigma} (u) D_{\mu\nu,\rho\sigma} (z,u)+
\right.$$
\be
\left.+\frac{i}{N_c-1}\sum_{i\neq j} \int \frac{dt_i}{2\mu_i}
ds^{(j)}_{\rho\sigma} (u) \sigma^{(i)}_{\mu\nu}
D_{\mu\nu,\rho\sigma}(z,u)\right\} \label{50}\ee

In (\ref{50}) we have defined as in Appendix $$
D_{\mu\nu,\rho\sigma} (z,u) =  \frac{g^2}{N_c} \lan tr F_{\mu\nu}
(z(t_i)) F_{\rho\sigma} (u) \ran = D(h) (\delta_{\mu\rho}
\delta_{\nu\sigma} -\delta_{\mu\sigma}\delta_{\nu\rho}) $$ $$
+\frac12 [\partial_{\mu} h_\rho \delta_{\nu\sigma} +perm.]D_1(h)
$$ and  $h_\mu=z_\mu(t_i) -u_\mu$.

The Dirac structure of the exponent in (\ref{50}) is a sum, which
can be written with notations from (\ref{44}) as \be V^{(so)}
=\sum^3_{i=1} V_i^{(so,diag)} (\hat 1_i\times \hat 1_{jk}) +
V_i^{(so,nondiag)} (\hat \rho^{(i)}_1\times \hat 1_{jk}), ~~ i\neq
j,k.\label{51} \ee

 In the $4\times
4$ matrix $\sigma^{(i)}_{\mu\nu} $ we first consider the diagonal
part. Repeating all the steps leading to (\ref{18}) and
(\ref{23}), (\ref{24}) one has the same expressions with the
replacement $m_i\to \mu_i~~, i=1,2,3,$ namely
\be
V_{{\SD}}^{(so~diag)}=\left\{ \sum^3_{i=1}
\frac{\vesig^{(i)}\veL^{(i)}}{2\mu^2_i}\frac{1}{R^{(i)}}
\frac{dV_1}{dR^{(i)}} + \frac{1}{N_c-1}
\sum_{i<j}\frac{\vesig^{(i)}\veL^{(i)}+\vesig^{(j)}\veL^{(i)}}{2\mu_i\mu_j}
\frac{1}{R^{(j)}} \frac{dV_2}{dR^{(j)}}\right\} .\label{52}\ee

Let us now consider the nondiagonal part in (\ref{51}), which can
be written as $$ \lan W_3\exp (g\sigma F)\ran_{(so,nondiag)} =
\exp \left\{ i\sum^3_{i=1}\int\frac{dt_i}{2\mu_i} \sigma _k^{(i)}
(D^{(ii)}_{k4,l4}(z,u) ds^{(i)}_{l4} (u)\right. $$
\be
+D_{k4,mn}(zu) ds^{(i)}_{nm} (u))+ i\sum_{i\neq j}
\int\frac{dt_i}{2\mu_i} \sigma_k^{(i)}\left(D^{(ij)}_{k4,l4} (z,u)
ds^{(j)}_{l4} (u)+ D^{(ij)}_{k4nm} ds^{(j)}_{nm} (u)
\right)_{ij}\label{53}\ee

 Now
taking into account (\ref{13}), (\ref{14}) one has  writing
(\ref{53}) in the form
 \be\lan W_3\exp (g\sigma
F)\ran_{(so,nondiag)}=\exp \left[-T\left\{
\sum^3_{i=1}V^{(ii)}_{so, nondiag)} + \sum^3_{i<j=1}V^{(ij)}_{(so,
nondiag)} \right]\right\}\label{54}\ee one finds from (\ref{53}),
replacing $D_{\mu\nu,\rho\sigma}$ from  (A1) and using (\ref{13}),
(\ref{14})
\be
V^{(ii)}_{(so, nondiag)} = \Delta^{(ii)}_{EE} +\tilde
\Delta^{(ii)}_{EE} +\Delta^{(ii)}_{EB}\label{55}\ee and  similarly
for $V^{(ij)}_{(so, nondiag)}$. For terms on the r.h.s. of
(\ref{55}) one obtains $$ \Delta^{(ii)}_{EE}=-i \frac{\vesig
\veR^{(i)}}{2\mu_iR^{(i)}}\Lambda^{(ii)}, $$ $$\Lambda^{(ii)}
=\int^{R^{(i)}}_0 d\nu du (D(\nu, u)+ D_1+\nu^2\frac{\partial
D_1}{\partial\nu^2}) $$ $$\tilde \Delta^{(ii)}_{EE} =-i \int
(\vesig \veu) (\veR^{(i)}\veu) \frac{ \partial D_1}{\partial u^2}
(\nu, u) d\nu d\beta R^{(i)}, ~~\veu=\veR^{(i)} \beta; $$
\be
\Delta^{(ii)}_{EB} =\frac{i}{2\mu^3_i}\int^\infty_{-\infty} \nu^2
d\nu \frac{\partial D_1(\nu_1, h)}{\partial \nu^2} \int^1_0
\beta^{(i)} d\beta^{(i)} (\veL^{(i)} \times \vesig^{(i)})
\vep^{(i)}.\label{56}\ee

Similarly for $\Delta^{(ij)}$ one obtains from the second term on
the r.h.s. of (\ref{53})
 $$ \Delta^{(ii)}_{EE}=-i \frac{\vesig
\veR^{(j)}}{2\mu_iR^{(j)}}\frac{\Lambda^{(ij)}}{(N_c-1)}, $$
$$\Lambda^{(ij)} = R^{(i)}\int^1_0 d\beta^{(j)}\int^\infty_{-\infty}
 d\nu  (D(\nu,r^{(ij)})+D_1+\nu^2 \frac{\partial
D_1}{\partial\nu^2}) $$ $$\tilde \Delta^{(ii)}_{EE} =-i \int
\frac{(\vesig \ver^{(ij)}) (\veR^{(j)}\ver^{ij)})}{2\mu_i(N_c-1)}
d\nu \int^1_0 d\beta^{(j)} \frac{\partial D_1(\nu,
r^{(ij)})}{\partial (r^{(ij)})^2}$$
\be
\Delta^{(ij)}_{EB} =\frac{i}{2\mu_i\mu_j} \int^\infty_{-\infty}
\nu^2 d\nu \frac{\partial D_1(\nu, r^{(ij)})}{\partial \nu^2}
\int^1_0 \beta^{(j)} d\beta^{(j)} (\veL^{(j)} \times \vesig^{(j)})
(\frac{\vep^{(i)}}{\mu_i} +\beta^{(j)}
\frac{\vep^{(j)}}{\mu_j}).\label{57}\ee

Here we have defined $\ver^{(ij)} = \veR^{(i)}-\beta^{(j)}
\veR^{(j)}$.

\section{Discussion}.

Let us now discuss the results obtained in the paper. For the
heavy-quark case the nonperturbative dependent potential is given
in (\ref{8}), and the perturbative part in (\ref{42}), so that the
total SD potential is
\be
V_{\rm SD}^{(total)} = V_{\rm SD}^{(nonpert)} + V_{\rm
SD}^{(pert)}.\label{58}\ee

Perturbative part agrees with that obtained long ago in \cite{7}
and repeated in many subsequent papers. However in \cite{8} the
term $V^{(\varepsilon, pert)}$ is taken twice as big as (\ref{38})
(or corresponding term in \cite{7}). A possible modification for
light quarks, which can produce this increase is discussed later.

The nonperturbative part $V_{\rm SD}^{(nonpert)}$ (\ref{8})
consists of six terms, which were never fully written before.

Only asymptotics at large distances of the first term in (\ref{8})
has  been written before in \cite{27} and later in \cite{26}, one
can find it from (\ref{8})
\be
 V^{(nonpert)} (R^{(i)}\to \infty) =-\frac{\vesig^{(i)}\veL^{(i)}
 \sigma}{4m^2_iR^{(i)}}.\label{59}\ee
 In \cite{8} were instead postulated the pairwise
 nonperturbative spin-orbit forces, which contradict expressions
 derived in this paper and in \cite{26}. All other terms,
 proportional to $V'_2, V_3, V_4$ and $V_5$ have never been
 written for the 3q case, while for the  $q\bar q$ case the
 corresponding terms (except for $V_5$) have been written in
 \cite{18}, \cite{19}. The term $V_5$ which has no counterpart in
 the $q\bar q$ case, is completely new, and its physical
 implication is still unclear.

 We now turn to be light quark case. Here the total SD "potential"
 is in general a sum of product of $(4\times 4)$ matrices, which
 can be written as
 \be
 \hat V_{\rm SD}^{(light~quarks)} = \sum_{i,j} (V_{diag}^{(ij)}
 \hat 1_i\times \hat 1_j+\hat V_{nondiag.}^{(ij)})\label{60}\ee
 where $\hat V^{(ij)}_{nondiag}$   contains terms like $\hat
 1_i\times \hat \rho_{1j},\hat \rho_{1i}\times \hat 1_j, \hat
 \rho_{1i} \times \hat \rho_{1j}$, and $\rho_1=\left (
 \begin{array}{ll}
 0&\hat 1\\
 \hat 1& 0\end{array} \right)$, where each  entry in $\rho_1$, is
 $2\times 2$ unit matrix.

 Now for $V_{diag}^{(ij)}$ one has
 \be
 V^{(ij)}_{diag} =V^{(\sigma\sigma)}_{\rm SD} (BB) + V_{\rm
 SD}^{(so, diag)}\label{61}\ee
 where the first term on r.h.s. of (\ref{61}) is given in
 (\ref{45}) and the second in (\ref{52}). One can see in these
 expressions for $V_{diag}^{(ij)}$ the same terms as in (\ref{8})
 with exchange $m_i\to \mu_i$ except for the spin-orbit term
 proportional to $\frac{d\varepsilon}{dR}$. Before discussing two
 different strategies for obtaining this last term, let us look at
 the general structure of (\ref{60}). It has the described above
 matrix form and depends on einbein fields $\mu_i, ~~i=1,2,3$. The
 latter have been defined previously in \cite{10,17,20, 23, 28} as
 scalars, $2\mu_i=\frac{dz_i(\tau)}{d\tau}$, and are assumed to be
 found from the stationary point  equation in the path-integral
 form of the meson Green's function, or from the stationary point
 of the Hamiltonian. Now the spin-independent part of Hamiltonian
 is a unit matrix and hence can produce scalar stationary values
 for $\mu_i$. The situation changes however if one tries to
 incorporate also the SD part of Hamiltonian in the stationary
 point equation for $\mu$, since it would require $\mu$ to have a
 matrix form similar to that of $\hat V_{\rm SD}$.

 This is possible in the generalized form of the FFFR, which is
 now under investigation, but in the present form the only
 possible way of treatment the SD part of Hamiltonian is to
 consider it as a perturbation. For light quarks it is not an
 expansion in $1/\mu_i$, and the whole expression (\ref{60}) is
 obtained with the only and numerically good approximation --
 keeping the bilocal (Gaussian) correlator, neglecting all higher
 ones.

 As it was shown in meson case \cite{20,28}, this perturbation
 procedure works well even for lowest mesons, where SD corrections
 produce up to around 15\% of the total mass (a similar situation
 holds true in earlier quark model calculations with fixed and
 prescribed constituent masses, see e.g. \cite{8}).  For heavier
 meson and baryon states the masses $\mu_i$ grow rapidly with
 quantum numbers \cite{20} and validity of perturbative treatment
 of SD terms becomes even better established. In what follows we
 describe a perturbative procedure of treating the nondiagonal terms.

 To this end we must remember (as in point B of derivation in
 section 4) that nondiagonal terms are also present in the
 preexponential factor $(m-\hat D)$ in (\ref{1}). Consider the
 largest nondiagonal term $\Delta^{(ii)}_{EE} $ in (\ref{56}), and
 take for simplicity its asymptotic form
 \be
 V_{EE}--i\sum^3_{i=1}
 \frac{\veal^{(i)}\ven^{(i)}\sigma}{2\mu_i},~~
 \ven^{(i)}=\veR^{(i)}/R^{(i)}.\label{62}\ee

 One has (omitting index $(i)$ for simplicity)
 \be
 (m-\hat D)\exp(-V_{EE} T)\cong\left ( \begin{array}{ll}
 m+\mu,&-\vesig\vep\\
 \vesig\vep,& m=\mu\end{array} \right) (1+i\frac{\sigma}{2\mu}
 \veal \ven +...T).\label{63}\ee

 Comparing with the leading term, given by the upper left corner,
 one normalizes $(m-\hat D)$ by extracting the factor $(m+\mu)$ and
 thus obtains
 \be
 Eq.(63) =(1-V_{\rm SD}^{(\varepsilon)}T+...),~~
 V_{\rm SD}^{(\varepsilon)}
 =\sum^3_{i=1}\frac{\sigma\vesig^{(i)}\veL^{(i)}}{2\mu_i(m_i+\mu_i)
 R^{(i)}}.\label{64}\ee

 In the heavy quark limit, $\mu_i\approx m_i$ and (\ref{61})
 coincides with the term proportional to
 $\frac{1}{R}\frac{d\varepsilon}{dR}$ in (\ref{8}). For light
 quarks , when $\mu_i\gg m_i$ however one has twice as large
 coefficient in(\ref{61}) which coincides with the heavy-quark
 expansion with the light quark expression. Thus our total
 expression for SD potential treated as perturbation is a $2\times
 2$ matrix
 \be
 \tilde V_{\rm SD}^{(light~quarks)} =\sum_{i,j} V_{diag}^{(ij)} +
 V_{\rm SD}^{(\varepsilon)}\label{65}\ee
 where $V^{(ij)}_{diag}$ is given in (\ref{61}), (\ref{45}),
 (\ref{52}) and $V_{\rm SD}^{(\varepsilon)}$ is given in
 (\ref{64}), with general form obtained by replacing $\sigma \to
 \frac{\partial \varepsilon}{\partial R}$. Now one can see from
 these expressions that we have a full correspondence between terms
 in (\ref{8}) and in (\ref{65}), where each term in (\ref{65}) is
 obtained from the corresponding one in (\ref{8}) by  replacement
 $m_i\to \mu_i$, except for the term with
 $\frac{d\varepsilon}{dR}$, where one replaces $2m^2_i\to
 \mu_i(\mu_i+m_i)$.

 \section{Concluding remarks}

 We have obtained all perturbative and nonperturbative
 spin-dependent terms in the 3q system in the approximation when
 lowest (bilocal) field correlator is retained in Wilson loop. The
 analogous procedure for mesons in \cite{18, 19} yielded SD
 potentials satisfying Gromes relation \cite{24}, with correct
 asymptotics at large distances of Thomas precession type.For
  the 3q system we also get this asymptotics for spin-orbit terms
  in the form of a sum of one-body Thomas terms, in agreement
  with earlier results in \cite{26}.
 All other nonperturbative terms and exact nonasymptotic form of
 Thomas terms are new. The signs of perturbative and
 nonperturbative spin-orbit terms are different and one may expect
 some cancellation, which should be checked in exact calculations
 of baryon spectra with spin splittings.
 All nonperturbative SD terms in (\ref{8}) except for $V_5$ have the
 structure similar to that of the $Q\bar{Q}$ case,considered in
  \cite{19},except that spin-orbit terms are of one-body rather
  than the two-body character.The new term $V_5$ (\ref{24}) does not
   have a $Q\bar{Q}$ analog,and after averaging over coordinates
 has a structure similar to two-body spin-orbit force.

 The large $N_c$ structure of SD interaction can be clearly seen
 from explicit expressions and may be represented as leading
 $(O(N_c^0))$ terms of one-body spin-orbit interaction, when both
 fields in the field correlator are on the same sheet of the
 3-sheet surface, and suppressed $(O(N_c^{-1}))$ terms of
 spin-spin interactions and spin-orbit from two different sheets.
 Hence the 3q dynamics in the large $N_c$ limit reduces to the
 uncorrelated motion of $N_c$ quarks around a common center
 (string junction), which can be taken as infinitely heavy.

 The general structure of the SD potential (\ref{8}) at large $N_c$
 is in agreement with the classification done in \cite{29}
 where the unsuppressed at large $N_c$ terms are one-body spin-orbit
 potentials,while two-body spin-dependent terms are $1/N_c$
 suppressed.In addition in \cite{29} appear also spin-flavour terms
 which can be associated with with pion and kaon exchange forces.
 The latter were not considered in the present paper,but can be
 easily included in the same formalism,using the new chiral
 Lagrangian derived in \cite{30}.It is shown there that in the
 $q-string-\bar{q}$ system pions are emitted by quarks with the
 known amplitude,so that the pion-exchange force can be predicted
 unambiguosly and added to the those obtained in the present work.
 This would complete the overall picture of SD forces in baryon.
 The author is grateful to J.Goity for useful discussion,remarks and
 suggestions.

 This work was supported by DOE contract DE-ACOS-84ER 40150 under
 which SURA operates the Thomas Jefferson National Accelerator
 Facility.

 \setcounter{equation}{0}
\renewcommand{\theequation}{A.\arabic{equation}}

\begin{center}
{\bf Appendix A}\\

{\large Field correlators}
\end{center}

>From general definitions of $g^2\lan  F_{\mu\nu}
F_{\lambda\sigma}\ran$ through $D, D_1$ in \cite{11} one gets $$
\frac{g^2}{N_c} tr \lan F_{\mu\nu} (x) \Phi(x, y) F_{\rho\sigma}
(y) \Phi(y,x) \ran = (\delta_{\mu\rho} \delta_{\nu\sigma}
-\delta_{\mu\sigma} \delta_{\nu\rho} ) D(z)+ $$
\be
+ \frac12[\partial_\mu z_\rho \delta_{\nu\rho} + perm. ] D_1(z)
\label{A.1} \ee $$
 \frac{g^2}{N_c} tr \lan B_i (x) \Phi(x, y) B_j
(y) \Phi(y,x) \ran
 = (\delta_{ij}(D(z) + D_1(z) +\vez^2
 \frac{\partial D_1}{\partial z^2}]-
$$
\be
-z_iz_j \frac{\partial D_1}{\partial z^2}\label{A.2}\ee $$
 \frac{g^2}{N_c} tr \lan E_i (x) \Phi(x, y) E_j
(y) \Phi(y,x) \ran
 = (\delta_{ij}(D(z) + D_1(z) +z^2_4
 \frac{\partial D_1}{\partial z^2}]+
$$
\be
+z_iz_j \frac{\partial D_1}{\partial z^2}\label{A.3}\ee \be
 \frac{g^2}{N_c} tr \lan B_i (x) \Phi(x, y) E_j
(y) \Phi(y,x) \ran
 = e_{ijk} z_4z_k
 \frac{\partial D_1}{\partial z^2}
\label{A.4} \ee where we have defined $$ z_\mu=x_\mu-y_\mu,~~
\mu=1,2,3,4, ~~ \Phi(x,y) =P\exp ig \int^x_y A_\mu(u) du_\mu $$

\end{document}